# Prediction of Residual Stress Due to Early Age Behaviour of Massive Concrete Structures:
# On Site Experiments and Macroscopic Modelling

**Jihad Zreiki** [1]
**Vincent Lamour** [2]
**Mohend Chaouche** [3]
**Micheline Moranville** [4]

T 11

**ABSTRACT**

Early age behaviour of concrete is based on complex multi-physical and multiscale phenomena. The predication of both cracking risk and residual stresses in hardened concrete structures is still a challenging task. We propose in this paper a practical method to characterize in the construction site the material parameters and to identify a macroscopic model from simple tests. We propose for instance to use a restrained shrinkage ring test to identify a basic early age creep model based on a simple ageing visco-elastic Kelvin model. The strain data obtained from this test can be treated through an early age finite element incremental procedure such that the fitting parameters of the creep law can be quickly identified. The others properties of concrete have been measured at different ages (elastic properties, hydration kinetics, and coefficient of thermal expansion). From the identified early age model, we computed the temperature rise and the stress development in a non reinforced concrete stress for nuclear waste. The good agreement between in-situ measurement and predicted behaviour allowed us to validate our approach.

[1]  LMT-Cachan Laboratory of Mechanic and Technology, Sector Civil Engrg, Cachan, France 94230, Phone +33 (0)1 47 40 22 38, Fax +33 (0)1 47 40 22 40, zreiki@lmt.ens-cachan.fr
[2]  LMT-Cachan Laboratory of Mechanic and Technology, Sector Civil Engrg, Cachan, France 94230, Phone +33 (0)1 47 40 22 38, Fax +33 (0)1 47 40 22 40, lamour@lmt.ens-cachan.fr
[3]  LMT-Cachan Laboratory of Mechanic and Technology, Sector Civil Engrg, Cachan, France 94230, Phone +33 (0)1 47 40 22 38, Fax +33 (0)1 47 40 22 40, chaouche@lmt.ens-cachan.fr
[4]  LMT-Cachan Laboratory of Mechanic and Technology, Sector Civil Engrg, Cachan, France 94230, Phone +33 (0)1 47 40 22 38, Fax +33 (0)1 47 40 22 40, moranville@lmt.ens-cachan.fr

**1 INTRODUCTION**

Early age behaviour of concrete is based on complex multi-physical and multiscale phenomena. The predication of the cracking risk and the residual stresses in the hardened concrete structures is still a challenging task. The use of the restrained shrinkage ring test to identify a basic early age creep model is based on a simple ageing visco-elastic Kelvin model.
We propose in this article to identify a simple macroscopic model with a reduced number of parameters. First we model the total behaviour of the structure in various configurations, and then specimens of building site are used to accomplish the identification.
The behaviour of the concrete at early age is generally described as visco-elastic. The rheological models based on Kelvin chains are used to model this behaviour [De Schutter 1996] and [Băzant *et al.* 1997], or they are based on Maxwell chains to take into account the humidity or temperature effects [Băzant & Chern 1985] and [Băzant and *al.*2004]. However, others models have used the Burger model [Hauggaard *et al.* 1999] and [Băzant et *al.*2004]. In section 2 we begin with the model used to describe elastic modulus of concrete at early age, and then we proposed a model based on the hydration degree. Section 3 focuses on the basic creep's and the function of creep, and then we describe the hydration kinetic. After having exposed the behavioural model of concrete at early age and having identified its parameters, a comparison is done between real measurements and outputs of the numerical model are detailed in section 4 before concluding.
The stake of such tool is to limit the early age cracking risk in order to increase the durability of the prefabricated products of concrete and to determine of the residual stresses of fabrication.

**2 EVOLUTION OF THE MODULE OF ELASTICITY**

There are several models describing the evolution of the Young modulus. Some of them require more tests to reproduce the material behaviour. Researches have shown that models using the Apparent Setting time will be more robust against effects of insufficiency in amount of experimental data [Larson & Jonasson 2003]. Larson proposed a Linear Logarithmic Model (LLM) for the creep of concrete, in which the Young modulus is given by equation 1:

$$E(t_0) = E_{ref}.\beta_E(t_0) \qquad (1)$$

Where: $E_{ref}$ is a reference value, which here is chosen as the Young modulus at 28 days age, and $\beta_E(t_0)$ is a parameter expressed as:

$$\beta_E(t_0) \begin{cases} 0 & t_0 < t_s \\ b_1.\log(t_0/t_s) & t_s \leq t_0 < t_s \\ b_1.\log(t_B/t_s) + b_2.\log(t_0/t_B) & t_B \leq t_0 < 28j \\ 1 & t_0 \geq 28j \end{cases} \qquad (2)$$

The traditional equations which describe the non-aging Young modulus, based on the compressive strength at age of 28 days, as in [Table 1], [Takács 2002].

**Table 1.** Formula for the Young modulus at age of 28 days.

| Code | Formula de $E_c$ (MPa) |
|---|---|
| CEP-FIP model code 1990 | $E_c = 9980(f_{cm})^{1/3}$ |
| Euro code 2 | $E_c = 9500(f_{cm})^{1/3}$ |
| ACI 318 | $E_c = 4733(f_{cm})^{0.5}$ |

We find in ACI-209R-82 a recommendation for the prediction of creep and shrinkage, and they gave a relation for calculates the aging Young modulus [Rajeev *et al.* 2007] Equation 3.

$$E_c(t) = 0{,}04326\sqrt{\rho^3 . f_{cm}\left(\frac{t}{a+b.t}\right)} \qquad (3)$$

Where: a, b: constants depend on cement type and treatment conditions; $E_c(t)$: The Young modulus of concrete at age t; $\rho$: The density of the concrete (kg/m$^3$).

The Young modulus given in CEB-FIP for a concrete of strength (12-80 MPa) at 28 days by the following form equation:

$$E_c(t_0) = E_c . \sqrt{\exp\left\{s\left(1 - \sqrt{\frac{18}{t_0}}\right)\right\}} \qquad (4)$$

Where: $E_c(t_0)$ The Young modulus of concrete at age $t_0$; c: Constant dependent on the type of cement.

In reality the Young modulus (E) of concrete evolves with time according to a chemical reaction of cement hydration [Ignacio & Băzant 1993] and [De Schutter 1999].

The Young modulus is given by De Schutter according to the degree of hydration for a concrete (CEM III/B 32.5) [De Schutter 1999].

$$E_0(\chi) = 37000 . \left(\frac{\chi - 0.25}{1 - 0.25}\right)^{0.62} MPa \qquad (5)$$

$\chi$ : The degree of hydration.

You can see a relation which depends on time, in which the Young modulus evolves in a potential way [Hauggaard *et al.* 1999].

$$E(t) = a.e^{-\left(\frac{b}{t}\right)^c} \qquad (6)$$

Where a, b, c are parameters depend on material, and t is the equivalent time defined at temperature 20±C.

Byfor used to define the Young modulus as function of degree of hydration [Eduardo *et al.* 2004] like:

$$E(\xi) = E_\infty . \frac{1 + 1{,}37(f_{c,\infty})^{2,204}}{1 + (f_c(\xi))^{2,204}} \left(\frac{f_c(\xi)}{f_{c,\infty}}\right) \qquad (7)$$

$E_\infty = E(\xi = 1)$ : The Young modulus at the end of hydration.

## 3 MACROSCOPIC MODEL OF CONCRETE AT EARLY AGE:

### 3.1 The creep's function:

The simplest and the oldest models of creep for concrete is the effective modulus method (EMM) [Băzant & Wittmann 1980], which consists only of a single linear elastic solution, as $E_{effective} = 1/J(t,t_0) = E(t_0)/(1+\phi(t,t_0))$, where J is the creep compliance function, $t_0$ is the age at loading, $E(t_0)$ is the initial Young modulus, and $\phi$ is the creep coefficient. The linear treatment of visco-elastic of ageing materials as the concrete can be characterised by the compliance function $J(t,t')$ or the relaxation function $R(t,t')$. $J(t,t')$ Presents the strain at time t caused by a constant stress applied at the time $t'$ [Ignacio & Băzant 1993]. For a stress different of 1, the strains are calculated by the following equation:

$$\varepsilon(t) = \Delta\sigma . J(t,t') \tag{8}$$

For $t = t_0$, we have $J(t_0, t_0) = 1/E(t_0)$. By applying the principle of superposition, we find the equation of Volterra which constitute the law of linear visco-elastic behaviour of aging material:

$$\gamma(t) = \int_{t'=0}^{t} J(t,t') d\sigma(t') \tag{9}$$

The use of Dirichlet series to expresser the function of creep $J(t,t')$, and the integral of the equation give more effective differential equations for the numerical solutions:

$$J(t,t') = \frac{1}{E_0(t')} + \sum_{\mu=1}^{n} \frac{1}{E_\mu(t)} \left(1 - e^{\frac{-E_\mu}{\eta_\mu}(t-t')}\right) \tag{10}$$

$J(t,t') = q_1 + C(t,t')$ Where $q_1$ is the instantaneous elastic strain. The equation of creep can be written like a sum of exponentials (i.e., Dirichlet series). It is possible to convert the equation (9) to linear differential equations. These equations can be simulated well by rheological models like Kelvin or Maxwell chains with an aging spring $E_\mu(t)$ and aging dashpot $\eta_\mu(t)$ [Ignacio & Băzant 1993]. There are several possibilities of arranging the spring and the dashpot, but it is shown that the behaviour of creep generally can be described by Maxwell or Kelvin chains [Băzant & Santosh 1989].
If we consider only one element of the chains of Kelvin (μ=1) 'Fig 1', we can write the differential equation:

$$E_1(\chi)\gamma + \eta_1(\chi)\dot{\gamma} = \sigma \tag{11}$$

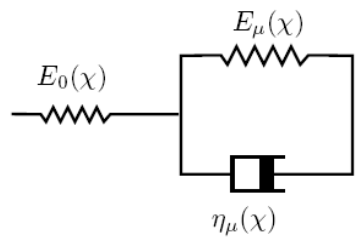

**Figure1.** The Rheological model of Kelvin

### 3.2 Thermal model:

We can write the equation of heat describing the thermal transfers in concrete at early age:

$$C(\chi)\frac{\partial T}{\partial t} = div(k(w,\chi).grad(T)) + Q_{hy}(\chi) - Q_{ev}(\chi) \tag{12}$$

Where C and K are the thermal capacity and conductivity, $Q_{hy}$ the heat provided by the hydration, $Q_{ev}$ the waste heat by evaporation in skin in the event of fast unmoulding in conditions of extreme drying.

### 3.3 Hydration kinetic:

With simplicity the kinetics of hydration of cement can be modelled by an equivalent chemical affinity according to Arrhenius's law:

$$A(\chi) = \dot{\chi} = \widetilde{A}(\chi).\exp\left(\frac{-E_a}{RT}\right) \qquad (13)$$

With: $\widetilde{A}(\chi)$ the chimical affinity equivalent standardized given for an isothermal reation of hydration.

The evolution and the heat of hydration of concrete can be followed by a quasi-adiabatic test on insulated cylindrical. We can determine the equivalent endogenous affinity and the evolution of the reaction advancement degree for various thermal paths.

We consider that the advancement of hydration reaction is stabilized at 28 days ($\chi = Q/Q_{28d} \approx 1$) 'Fig.2'.

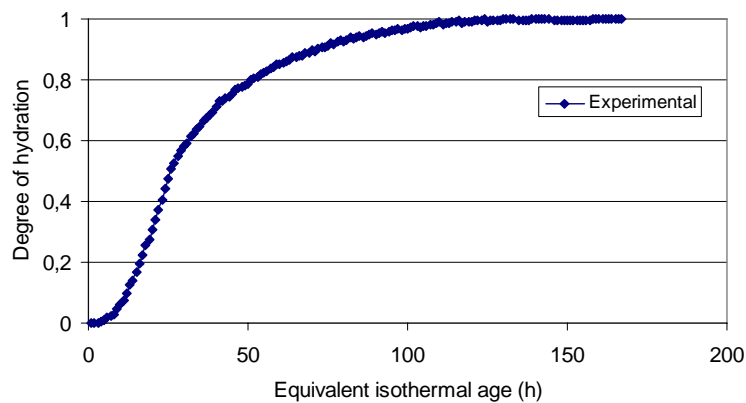

**Figure 2.** Evolution of degree of advance of reaction (activation energy = 46 KJ/mol)

### 3.4 Mechanic model:

We used a mechanical model based on the deformation partition principle:

$$\dot{\varepsilon} = \dot{\varepsilon}_e + \dot{\varepsilon}_c + \dot{\varepsilon}_T + \dot{\varepsilon}_{Sh} \qquad (15)$$

With: $\dot{\varepsilon}_T = \alpha.\dot{T}$ the thermal deformation, $\alpha$ the thermal dilation coefficient; $\dot{\varepsilon}_{Sh}$ the shrinkage's deformation, $\dot{\varepsilon}_e$ elastic deformation, $\dot{\varepsilon}_c$ creep's deformation equation (11).

## 4 IDENTIFICATION OF MODELE

### 4.1 Used Concrete

We used to identify the model a self compression fibred concrete with a low heat of hydration, the composition of this concrete is presented in the Table 2

### 4.1 Mechanical proprieties

To identify the creep model, we propose a shrinkage ring test. In fact the ring test is a simple test by which we measure the visco-elastic deformation in a ring of concrete. The geometry of the ring is presented in the 'Fig.3'.

**Table 2.** Formulation of concrete (W/C = 0.35).

| Materials | Concrete fractions [Kg/m$^3$] |
|---|---|
| Cement CEM V / A (S-V) 42.5 N CE PM ES CP1 | 454 |
| Silica fume | 45 |
| Calcareous sand 0/4mm | 984 |
| Calcareous aggregate | 672 |
| Affective water | 173 |
| Superplasticizer | 5.2 |
| Fibre (L30mm, Ø0,6mm) | 85 |

The section of the brass ring: 2*7cm, the section of concrete: 7*7cm, the outside diameter of concrete: 64 cm, and the inside diameter: 50cm, the inside diameter of ring: 46 cm. We measure orthoradial deformation by four gauges presented in 'Fig.5'. The rheological diagram of the ring test is represented on 'Fig 4' [Lamour & al 2007]. The result of the ring test allows us to determine the parameters of the creep model by an explicit diagram of integration. 'Fig 5'

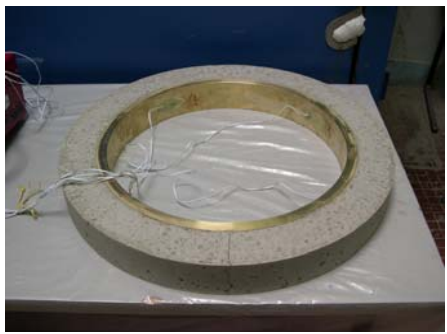
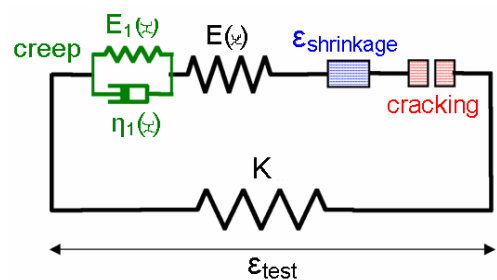

**Figure 3.** The ring test        **Figure 4.** Simplified 1D parallel model for the ring test.

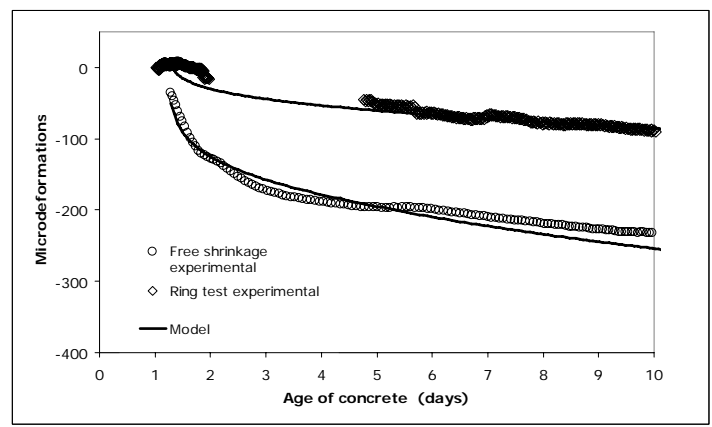

**Figure 5.** Simplified 1D parallel model for the ring test

The evolution of the elastic modulus was measured on specimen (7*7*28 cm ) by using the Ultrasound velocity measurement, where transversal wave velocities were measured, the elastic modulus (E) was determined by using wave propagation theory in homogeneous bodies.
We propose for fitting elastic modulus a model (see eq.5 )based on $\chi$ (hydration degree) like the model proposed by [De Schutter G. 1999].

$$\left\{ E(\chi) = E_{max} \left( \frac{\chi - \chi_0}{\chi_{max} - \chi_0} \right)^b \right\} \tag{16}$$

Where : $\chi$ the hdration degre; $\chi_{max} \approx 1$ the hydration degree at end of test; $\chi_0 = 0.25$ the hydration degree at unmoulding, $E_{max} = 35 GPa$ the elastic modulus at fin of essais, b = 0.5 coefficient . The experimental resalting and modele proposed showen in 'Fig 6 '

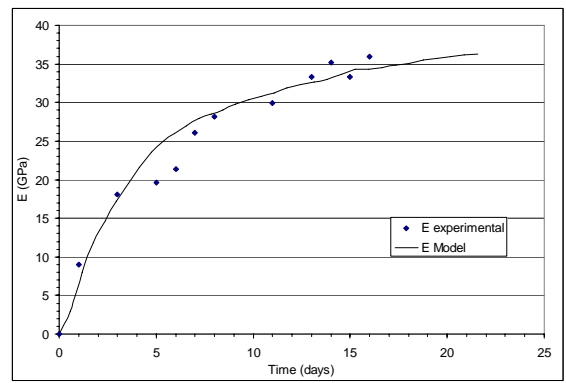

**Figure 6.** Elastic modulus

## 5 CALCULUS OF THE RESIDUAL CONSTRAINT ON REAL STRUCTURE

The studies structure corresponds to a piece of fibres massive concrete none armed including four reservations 'fig 7'. The measurement of the deformation and temperature is automated via an autonomous power station of acquisition.

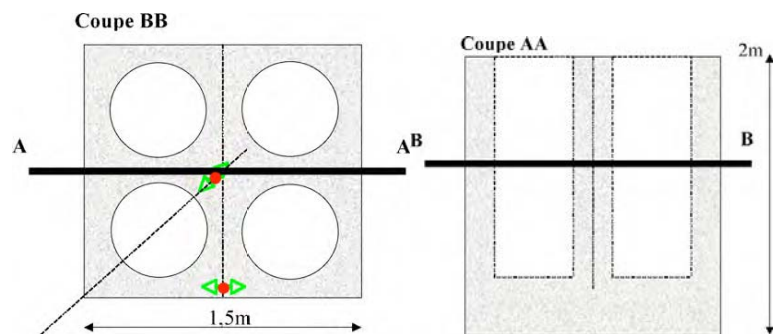

**Figure 7.** The piece of concrete with two vibrating cord sensor installation (red and green)

The model is implanted in code of element finis (CAST3M) developed by [CEA-France], the resolution of problem is done in an iterative way: for each step of time corresponding to an increment of the advance of the reaction of hydration in each element [Lamour & *al* 2007]. The deformations are calculated from an identified model on specimen, moreover the modelling of structure makes it possible to evaluate the mechanical constraints generated in the structure 'Fig 8'. We verify in particular that the generated constraints are lower than resistances in traction of the concrete.

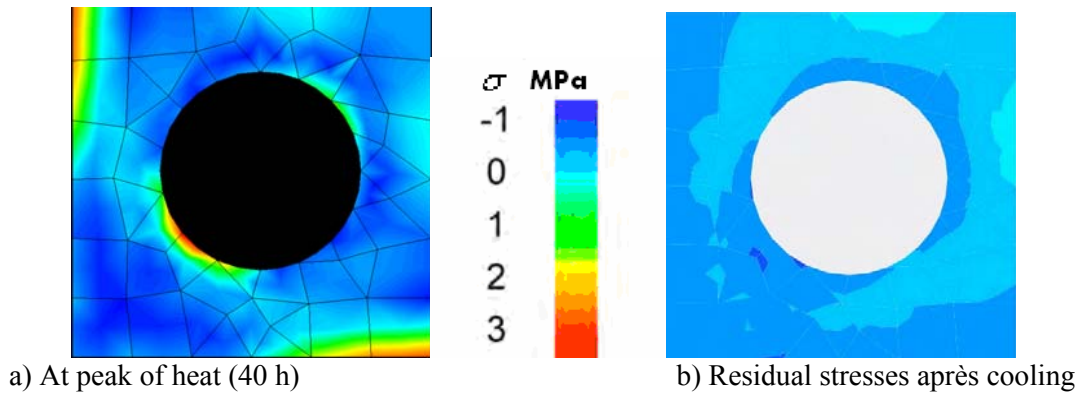

a) At peak of heat (40 h)  b) Residual stresses après cooling

**Figure 8.** Simulated cartography of the principal stress in a quarter of the piece
(The blue zones are a compression zones, the orange zones represent the zones at the risk with respect to cracking at early age).

## 6 CONCLUSIONS

We proposed in this paper a practical method to characterize the material parameters in the construction site and to identify a macroscopic model from simple tests. We proposed for instance to use the restrained shrinkage ring test to identify a basic early age creep model based on a simple ageing visco-elastic Kelvin model.

Model identification of elastic modulus based on hydration degree has given good resultants. The identification of creep's deformation was done by a rheological model in using the ring test.
This model makes it possible to quantify the risk of cracking and to evaluate the residual stresses of manufacture by taking into account the clean creep identified by the ring test.